\documentclass[aps,twocolumn,10pt,showpacs]{revtex4}
\usepackage{graphicx}
\usepackage{graphicx}		
\usepackage{subfigure}
\usepackage{enumerate}
\usepackage{color}
\usepackage{tabulary}
\usepackage{tabularx}
\usepackage{amssymb}	
\usepackage{amsfonts}		
\usepackage{amsmath}
\usepackage{subfig}
\usepackage{caption}		

\begin{document}

\title{Enhancing the spreading of quantum walks on star graphs by additional bonds}


\author{Anastasiia~Anishchenko, Alexander~Blumen and Oliver~M{\"u}lken}	


\affiliation{Theoretische Polymerphysik, Universit\"at Freiburg,
              Hermann-Herder-Stra\ss e 3, D-79104 Freiburg, Germany \\
              Tel.: +49-761-20397783\\
              \email{anastasiia.anishchenko@physik.uni-freiburg.de}           
         }

\date{\today}

\begin{abstract}
We study the dynamics of continuous-time quantum walks (CTQW) on networks
with highly degenerate eigenvalue spectra of the corresponding
connectivity matrices. In particular, we consider the two cases of a star
graph and of a complete graph, both having one highly degenerate
eigenvalue, while displaying different topologies.  
While the CTQW spreading over the network - in terms of the average probability to return or to 
stay at an initially excited node - is in
both cases very slow, also when compared to the corresponding classical
continuous-time random walk (CTRW), we show how the spreading is enhanced
by randomly adding bonds to the star graph or removing bonds from the complete graph. 
Then, the spreading of the excitations may become very fast, even outperforming the corresponding CTRW. 
Our numerical results suggest that the maximal spreading is
reached halfway between the star
graph and the complete graph.
We further show how this disorder-enhanced spreading is related to the networks' eigenvalues.
\end{abstract}


\maketitle

\section{Introduction}
\label{intro}
Transfer processes play a very important role in many physical,
chemical or biological instances, involving energy, mass or charge. Such transport processes strongly depend on the
topology of the system under study, as, for instance in simple crystals
\cite{ziman1979principles}, complex molecules \cite{kenkre1982exciton} or spin networks \cite{burgarth2008coupling}.

One often encounters regular networks, such as lattices or rings, where most or 
all of the nodes have the same degree. Examples are the star graphs (SG), and the complete graphs (CG), which have attracted lately much
attention in the field of quantum computation. Such studies focused on the quantum central limit theorem for continuous-time quantum
walks on SG \cite{salimi2009continuous}, the discrete-time
Grover walk on a SG with one loop \cite{machida2011discrete}, and quantum searches
on highly symmetric (complete) graphs \cite{PhysRevA.79.012323}. Exact analytical results were recently obtained for quantum walks on SG 
and on CG whose Hamiltonians have one highly degenerate eigenvalue. This,
as also discussed in what follows, may let the quantum walk remain localized at its starting node, 
which then renders the quantum spreading slow.  

Spreading of an excitation can be made faster by randomly adding (deleting) bonds to 
the SG (from the CG). Here we do not allow the additional bonds to connect one node with
itself and forbid double bonds between any pair of nodes. The inclusion of the additional bonds changes the 
structure of the network and, consequently, the eigenvalue set of the 
corresponding Hamiltonian. Furthermore,
destroying the original regularity of the SG (CG), the additional bonds create
``shortcuts`` in the graph, by which the spreading through CTQW may occur faster, as, for instance, found for small-world
networks \cite{mülken2007quantum}.

For modelling transport processes on networks one can use several approaches: in
quantum mechanics, the Hamiltonian is determined by the connectivity of
the system. For example, the dynamics of an electron in a crystal can be
described by the Bloch ansatz \cite{ziman1979principles}. This can be also
related to transport processes in polymers, where the connectivity plays
an important role in the dynamics \cite{doi1988theory}.

Classical transport processes modelled by continuous-time random walks
(CTRW) are described by a master equation, involving a transfer operator $\bf
T$ based on the topology of the system \cite{kenkre1982exciton}, \cite{van1992stochastic}, \cite{JKempe}. A quantum mechanical
analogue of CTRW, namely, one variant of the continuous-time quantum walk (CTQW), can be
introduced by using the transfer operator $\bf T$ in defining the Hamiltonian
$\bf H$ \cite{farhi1998quantum}. For simple lattices this is equivalent to a nearest-neighbor
hopping model \cite{farhi1998quantum}, \cite{mülken2006coherent}, \cite{mülken2005spacetime}, \cite{mülken2005asymmetries}. 
The transformation replaces the classical diffusion
process by a quantal propagation through the structure.
 
The paper is organized as follows. Section \MakeUppercase{\romannumeral 2} 
introduces the CTRW and CTQW concepts. Tools for determining their spreading are displayed in Section \MakeUppercase{\romannumeral 3}. 
Section \MakeUppercase{\romannumeral 4} focuses on CTRW and CTQW on SG and on CG, whereas the spreading over 
graphs intermediate between the SG and the CG is studied in Section \MakeUppercase{\romannumeral 5}. 
Section \MakeUppercase{\romannumeral 6} concludes with a summary of results.

\section{Quantum walks on networks}

In the following we focus on the dynamics of excitations over networks; these consist of $N$ nodes connected by bonds. 
The information about a network's topology, i.e., its connectivity, is stored in the $N\times N$
connectivity matrix $\bf A$, whose elements $A_{kj}$ are:
 \begin{equation}
\label{eq:conn_metrix}
A_{kj} =
\left\{ 
 \begin{array}{ll}
 f_j & \mbox {if $k = j$} \\
  -1 & \mbox {if $k$ and $j$ are directly connected} \\
 0 & \mbox {otherwise}.
 \end{array}
 \right.
\end{equation}
Here $f_j$ is the number of bonds emanating from $j$, also referred to as the degree of $j$. 
The connectivity matrix has the following properties:
\begin{enumerate}
 \item $\bf A$ is symmetric and real.
 \item All its eigenvalues $E_n$ are real and non-negative.
\end{enumerate}
If the network is simply connected, $\bf A$  has a single vanishing eigenvalue $E_{\rm {min}}=0$.
In order to model the dynamics in the two extreme cases of a purely coherent
and of a purely incoherent (diffusive) transport, we employ the concepts of CTQW
and of CTRW, respectively. In both cases, the dynamics is largely
influenced by the network's topology, i.e., by $\mathbf A$. 
Now, we depict an excitation localized at node
$j$ through the state $|j\rangle$; the ensemble of the $|j\rangle$ states forms an 
orthonormal basis set $\{|j\rangle, j=1,...,N\}$. 
In the incoherent CTRW case, the spreading occurs based on the
transition rates $\gamma_{kj}$, which denote the probability to go per unit time from $j$ to $k$. Assuming these rates to be the same
for all the nodes, i.e., taking $\gamma_{kj}\equiv \gamma$ (and we will set $\gamma=1$
without loss of generality), induces a simple relation between $\mathbf T$ and $\mathbf A$, namely, $\mathbf T =
-\mathbf A$ \cite{van1992stochastic}, \cite{farhi1998quantum}, \cite{mulken2011continuous}.

In the quantum case, the states $|j\rangle$ span the whole accessible
Hilbert space. The time evolution of an excitation initially placed at
node $|j\rangle$ is determined by the system's Hamiltonian $\bf H$; in the approach of Ref. \cite{farhi1998quantum} used here, the 
CTQW is directly related to $\bf T$ through $\bf H=-\bf T$ \cite{farhi1998quantum}, \cite{mulken2011continuous}. 
The classical and quantum mechanical probabilities to be in state $|k\rangle$ at 
time $t$ when starting at $t=0$ from state $|j\rangle$ are then:
\begin{eqnarray}
&& p_{k,j}(t)=\langle k|\exp{(\mathbf{T}t)}|j\rangle 
\label{eq:trans_prob_cl}\\ 
\mbox{and} \qquad && \pi_{k,j}(t)=\arrowvert\langle\
k|\exp{(-i\mathbf{H}t)}|\ j\rangle \arrowvert^2,
\label{eq:trans_prob}
\end{eqnarray}
respectively, where we set in Eq.~(\ref{eq:trans_prob}) $\hbar=1$. For $t=0$ both expressions read $p_{k,j}(0)=\pi_{k,j}(0)=\delta_{k,j}$, 
where $\delta_{k,j}$ is Dirac's delta-function.

To get from the formal Eqs. (\ref{eq:trans_prob_cl}) and (\ref{eq:trans_prob}) to the explicit solutions 
for a particular lattice one has to diagonalize $\mathbf T$ and $\mathbf H$. Given that here the 
CTQW Hamiltonian $\mathbf H$ is the negative of the CTRW transfer matrix $\mathbf T$, lets the
eigenvalues and eigenvectors of both operators be practically identical. The difference in the dynamics of CTQW and of CTRW is due to the
different functional forms of Eq.~(\ref{eq:trans_prob_cl}) and of Eq.~(\ref{eq:trans_prob}). 

\section{Classical and quantum spreading}

While one needs to calculate the eigenvectors
(which can become tedious for large networks) for making definite statements on the transition probabilities
$p_{k,j}(t)$ and $\pi_{k,j}(t)$, there are quantities - to be
defined below - which only depend on the eigenvalues, whose distribution, the density of states (DOS) or spectral density, 
is given by:
\begin{equation}
  \rho(E) =\frac{1}{N}\sum_{n=1}^{N}\delta(E-E_N).
\end{equation}
The DOS contains information about the system and allows to analyse several features which depend on the network's topology. 
For CTRW one such feature is
the classical average
probability to be (return or remain) at the initially excited node 
averaged over all possible initial nodes $j$ \cite{alexander1982density}, \cite{bray1988diffusion}:

\begin{equation}
 \label{eq:ARP}
  \bar p(t)=\frac{1}{N}\sum_{j}{p_{j,j}(t)}=\frac{1}{N}\sum_{n=1}^{N}e^{-E_nt}.
\end{equation}
Hence $\bar p(t)$ depends only on the eigenvalues $E_n$ of $\bf A$, but not on its
eigenvectors $|\phi_n\rangle$.

For CTQW, on the other hand, $\bar {\pi}(t)\equiv\frac{1}{N}\sum_{n=1}^{N}\pi_{j,j}(t)$ depends on the eigenvectors $|\phi_n\rangle$. 
However, one can define, in a
fashion similar to the classical $\bar p(t)$, the quantity \cite{mülken2006coherent}, \cite{mülken2006efficiency},  
\begin{equation}
\label{eq:ARP_LowerBound}
\bar\alpha(t)\equiv\frac{1}{N}\sum_{j=1}^{N}\alpha_{j,j}(t) = \frac{1}{N}
\sum_{n=1}^N e^{-iE_nt}. 
\end{equation}

Here $\bar\alpha(t)$ also depends only on the eigenvalues $E_n$. Furthermore, $\arrowvert\bar\alpha(t)\arrowvert^2$ obeys the Cauchy-Schwarz 
inequality \cite{mülken2006coherent}:

\begin{equation}
 \bar {\pi}(t)= \frac{1}{N}\sum_{j}\Big
\arrowvert{\alpha_{j,j}(t)}\Big\arrowvert^2\geq \Big
\arrowvert\bar\alpha(t)\Big \arrowvert^2.
\end{equation}

Hence $\arrowvert\bar\alpha(t)\arrowvert^2$ is a lower bound to $\bar {\pi}(t)$.
We will use Eqs.~(\ref{eq:ARP}) and (\ref{eq:ARP_LowerBound}) to assess the spreading \cite{mülken2006efficiency}. 
Now, classically, a quick decrease of $\bar p(t)$ means a quick increase in the
probability of finding the excitation away from the initially
excited node. We hence infer that the spreading is faster when $\bar p(t)$ decreases
more quickly. In the quantum case, because of the unitary
time evolution, $\bar \pi(t)$ and $\arrowvert\bar
\alpha (t)\arrowvert^2$ show oscillations. However, the overall (average) decay of $\bar p(t)$ and of $\arrowvert\bar
\alpha (t)\arrowvert^2$  can be used to infer the spreading, say, by focusing on the decay of the envelope of, 
in particular, $\arrowvert\bar
\alpha (t)\arrowvert^2$ \cite{mülken2006efficiency}. 

For regular $d-$dimensional networks these procedures to assess the spreading have been used and discussed in
\cite{mülken2006efficiency}. The result was that quantum walks over such networks appear to be faster than the classical ones: 
the envelope of $\arrowvert\bar \alpha (t)\arrowvert^2$ turns out to decay as $t^{-d}$, whereas $\bar p(t)$ decays as $t^{\frac{-d}{2}}$.
Evidently, other networks may (and do) behave differently. In such cases we will
use the long-time averages 
\begin{eqnarray}
&& P_{\rm RW} \equiv \lim_{T\to\infty} \frac{1}{T} \int\limits_0^T dt \ \bar
p(t) \\
\mbox{and} \qquad && P_{\rm QW} \equiv \lim_{T\to\infty} \frac{1}{T} \int\limits_0^T dt \ \arrowvert\bar
\alpha (t)\arrowvert^2
\label{eq:ARP-QW-lta}
\end{eqnarray} 
in order to extract a time-averaged global spreading measure.

For CTRW, due to the eigenvalue $E_1=E_{\rm {min}}=0$,  $\bar p(t)$ will eventually drop to the
equipartition value $1/N$, as one can see by rewriting Eq.~(\ref{eq:ARP}) as:
 \begin{equation}
 \label{eq:ARP_Effic}
  \bar p(t)=\frac{1}{N}+\frac{1}{N}\sum_{n=2}^{N}e^{-E_nt}.
\end{equation}
It hence follows that $\lim_{T\to\infty}\bar p(t)=1/N$ and also that $P_{\rm RW} = 1/N$. Thus, in the
incoherent case at long times both quantities do not depend on the topology of the considered network.
In the quantum case, $\arrowvert\bar \alpha
(t)\arrowvert^2$ oscillates. With Eqs.~(\ref{eq:ARP_LowerBound}) and (\ref{eq:ARP-QW-lta}) it follows:
\begin{eqnarray}
\label{eq:ARP_lowerbound_lta}
P_{\rm QW} &=& \lim_{T\to\infty} \frac{1}{T} \int\limits_0^T dt \nonumber
\sum_{n,n'} e^{-i(E_n-E_{n'})t} \\
&=& \frac{1}{N^2} \sum_{n,n'} \delta(E_n - E_{n'}) \nonumber
= \frac{1}{N} \sum_{n} \rho(E_n) \\
&=& \sum_{E_n} [\rho(E_n)]^2.  
\end{eqnarray}
In Eq.~(\ref{eq:ARP_lowerbound_lta}) $\delta(E_n - E_{n'})$ is unity if $E_n=E_{n'}$ and vanishes otherwise. 
Note that $\sum_{E_n} \rho(E_n) = 1$. Thus, the DOS completely determines
the long-time average of $\arrowvert\bar \alpha
(t)\arrowvert^2$. For instance, for a ring with an odd number of nodes all eigenvalues are non-degenerate, i. e. there are $N$ 
different eigenvalues, each with $\rho(E_n)=1/N$, and therefore in that case
$P_{\rm QW}^{\rm ring} = 1/N$. The DOS determines hence the CTQW spreading measure $P_{\rm QW}$.

We can rewrite Eqs.~(\ref{eq:ARP}) and (\ref{eq:ARP_LowerBound}) as
\begin{equation}
 \label{eq:ARP_eff}
  \bar p(t)=\sum_{E_n}\rho({E_n})e^{-E_nt}
\end{equation}
and
\begin{equation}
\label{eq:ARP_LowerBound_eff}
 \bar {\pi}(t) \geq\Big \arrowvert\sum_{E_n}\rho({E_n})e^{-iE_nt}\Big \arrowvert^2=|\bar\alpha(t)|^2,
\end{equation}
respectively. In general, for a single highly degenerate eigenvalue $E_m$, with 
$\rho({E_m})$ of order $\mathcal O(1)$, while the other eigenvalues have a DOS at most of order $\mathcal O(1/N)$, the
average transition amplitude is \cite{mülken2006efficiency}:
\begin{equation}
 \bar {\alpha}(t) =\Big \{\rho({E_m})e^{-iE_mt}+\sum_{E_n\neq E_m}\rho({E_n})e^{-iE_nt}\Big \},
\label{eq:av_trans_ampl}
\end{equation}
from which one gets to order $\mathcal{O}(1/{N^2})$ the approximate expression:
\begin{equation}
\label{eq:one_highly}
 |\bar\alpha(t)|^2 \approx\rho({E_m})\Big \{\rho({E_m})+\sum_{E_n\neq E_m}\rho({E_n})\cos [(E_m-E_n)t]\Big \}.
\end{equation}
Equation (\ref{eq:one_highly}) shows that for highly degenerate
eigenvalues the lower bound $|\bar\alpha(t)|^2$ will not decay to zero, but will 
oscillate about a finite value. This means that for CTQW there
is a high probability to remain or to return to the initially excited
node. This leads to a slow average spreading for CTQW on such networks. 
However, for lattices of high dimensions, there is a significant probability that a walker never returns to the origin, 
thus, the CTQW can get transient \cite{darázs2010pólya}.
In the following we will focus on two particular examples of such networks
and on how to overcome the slow spreading.

\section{Star graph and complete graph}

\begin{figure}[h]
\centering
\includegraphics[width=0.4\columnwidth]{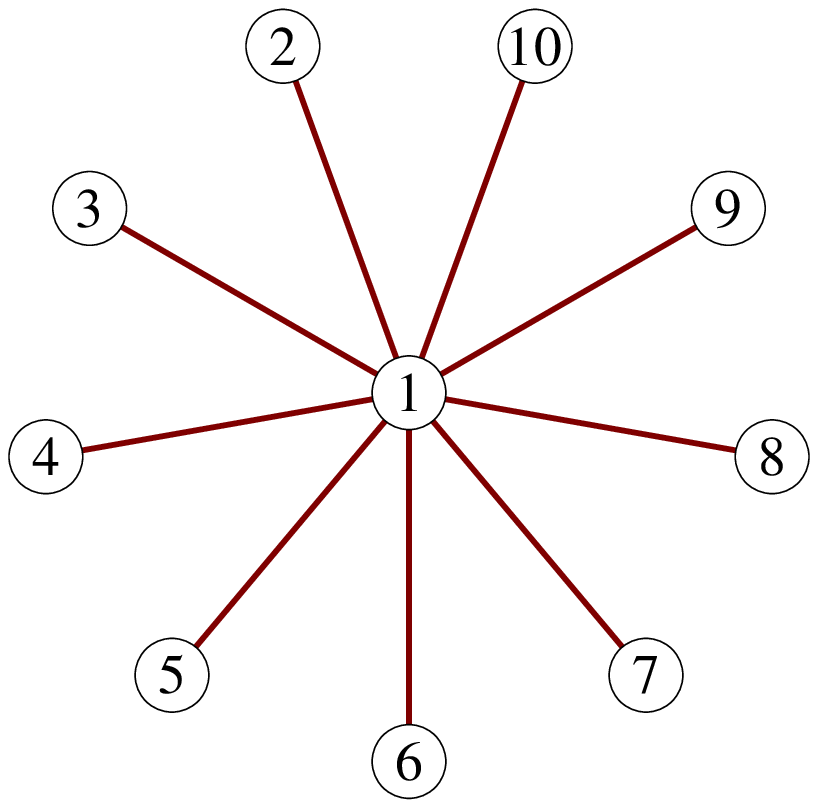}
\includegraphics[width=0.4\columnwidth]{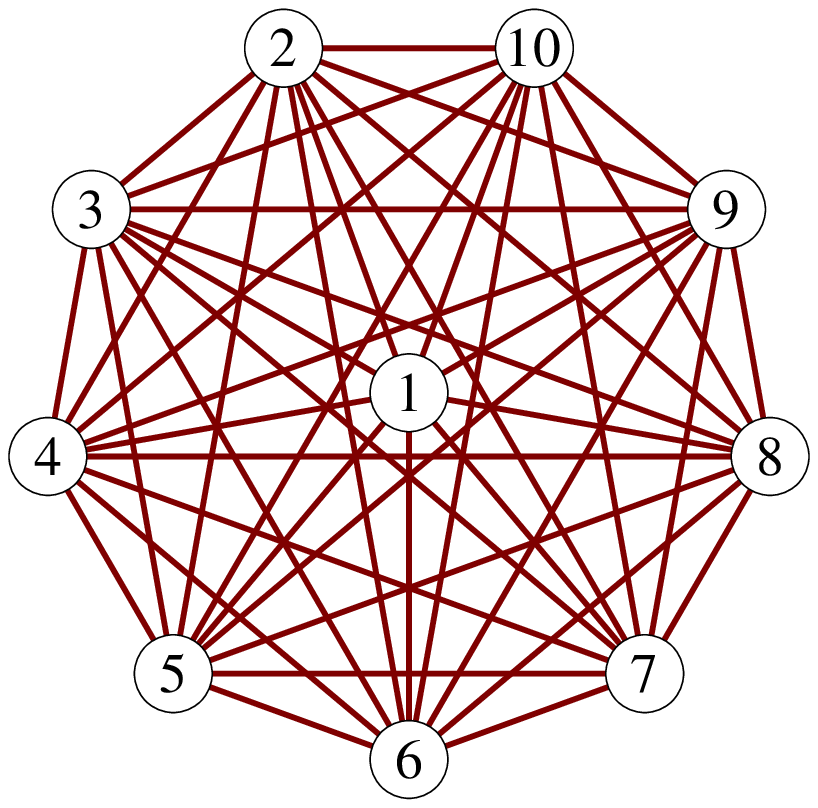}
\caption[StarGraph]{SG and CG of size $N=10$}
\label{fig:star_graph_0}
\end{figure}

There exist many networks with highly degenerate eigenvalues. Here, we 
focus on two specific networks which have both one highly
degenerate eigenvalue but vastly different topologies: On the one hand we consider
the SG, which is a network consisting of $N$ nodes with a central
node connected to $N-1$ leaf nodes, which are not connected to each other
(see Fig.~\ref{fig:star_graph_0}). Therefore, the central node has
degree $N-1$, and each of the leaf nodes has degree $1$. On the other hand we
consider the CG, where each of the $N$ nodes is
connected to all other $N-1$ nodes (see Fig.~\ref{fig:star_graph_0}).
Both networks are very ``ordered'', in the sense that there is an ``exchange''
symmetry, meaning that exchanging the positions of any pair of nodes
(except for the central node of the SG) leaves the network
invariant.

The connectivity matrices of the two graphs read
\begin{equation}
\label{eqn:star_matrix}
\bf A_{\rm SG} = \left(
\begin{array}{lllll}
N-1 & -1& \cdots & \cdots& -1\\
-1 & 1& 0& \cdots& 0\\
\vdots& 0& \ddots &  & \vdots \\
\vdots& \vdots &  & \ddots & 0\\
-1 & 0& \cdots& 0 & 1
\end{array}
\right)
\end{equation}
and
\begin{equation}
\label{eqn:complete_matrix}
\bf A_{\rm CG} = \left(
\begin{array}{lllll}
N-1 & -1& \cdots & \cdots& -1\\
-1 & N-1 & -1& \cdots& -1\\
\vdots& -1 & \ddots &  & \vdots \\
\vdots& \vdots &  & \ddots & -1\\
-1 & -1& \cdots& -1 & N-1
\end{array}
\right).
\end{equation}

Consequently, the Hamiltonians can be written as
\begin{equation}
\label{eq:star_hamiltonian}
 \mathbf{H}_{\rm SG} = (N-1)| 1\rangle \langle 1 | + \sum_{j=2}^{N} {(| j\rangle \langle j | - |1\rangle \langle j |- | j\rangle \langle 1 |)}
\end{equation}
and
\begin{equation}
\label{eq:complete_hamiltonian}
 \mathbf{H}_{\rm CG} = (N-1)\sum_{j} {(| j\rangle \langle j |)}-\sum_{j\neq i} {(| j\rangle \langle i |)}.
\end{equation}

The eigenvalues of both graphs can be calculated
analytically. The SG of size $N$ has only three distinct eigenvalues
\cite{mülken2006efficiency}: $E_1 = 0$, $E_2  = \dots = E_{N-1} = 1$,
and $E_N = N$ and their degeneracies are $D(0)=1$, $D(1)=N-2$, and $D(N)=1$, respectively. 
The CG of size $N$ has only two distinct
eigenvalues \cite{xu2009exact}, namely $E_1 = 0$ and $E_2= \dots =E_N = N$, with
degeneracies $D(0)=1$ and $D(N)=N-1$. 

The dynamics of CTQW and CTRW on SG and on CG has been studied in
\cite{xu2009exact}, where exact analytical results for the transition and
the return probabilities have been obtained. 
It has been shown that when the central node in the SG is initially excited, the spreading turns out to be equivalent to that over 
a CG of the same size.

Using Eq.~(\ref{eq:ARP_eff}) and (\ref{eq:ARP_LowerBound_eff}) one obtains the following analytical expressions
for $\bar {p}(t)$ and for $|\bar\alpha(t)|^2$ \cite{mulken2011continuous} for the SG :
\begin{equation}
\label{eq:star_eff_clas}
 \bar {p}(t)=\frac{1}{N}\Big \{1+(N-2)e^{-t}+e^{-(N-2)t}\Big \}
\end{equation}
and
\begin{equation}
\label{eq:star_eff_quantum}
|\bar\alpha(t)|^2=\frac{1}{N^2}\Big \arrowvert 1+(N-2)e^{-it}+e^{-i(N-2)t}\Big \arrowvert ^2.
\end{equation}

For the CG the analytical expressions for $\bar {p}(t)$ and $|\bar\alpha(t)|^2$ read \cite{mulken2011continuous}:

\begin{equation}
\label{eq:compl_eff_clas}
 \bar {p}(t)=\frac{1}{N}\Big \{1+(N-1)e^{-Nt}\Big \}
\end{equation}
and
\begin{equation}
\label{eq:compl_eff_quantum}
|\bar\alpha(t)|^2=\frac{1}{N^2}\Big \arrowvert 1+(N-1)e^{-iNt}\Big
\arrowvert ^2.
\end{equation}

It is a simple matter to calculate the long-time averages of the expressions above. 
For Eqs.~(\ref{eq:star_eff_clas}) and (\ref{eq:compl_eff_clas}) the CTRW long-time averages are ${P}_{\rm RW} = 1/N$ 
both for the SG and for the CG. In the quantum case we use $D(E_n)=N\rho(E_n)$ in Eq.~(\ref{eq:ARP_lowerbound_lta}). Such that, 
using now the explicit eigenvalues and their degeneracies listed above, we obtain:

\begin{eqnarray}
\label{eq:ARP_lta_SG}
&& P_{\rm QW}^{\rm SG} = \frac{N^2-4N+6}{N^2}\geq\frac{1}{N} \\
\mbox{and} \quad
\label{eq:ARP_lta_CG}
&& P_{\rm QW}^{\rm CG} = \frac{N^2-2N+2}{N^2}\geq\frac{1}{N}.
\end{eqnarray}

We conclude that a quantum walker on such networks in particular for large $N$, 
has a large probability to stay or to return to the initially excited
node, much larger than for a classical walker. We may even see this as a sign that the quantum spreading 
is slower than its classical analogue.
Clearly, this is
due to the fact that one eigenvalue of $\mathbf H$ is in both cases highly degenerate. 
One may compare Eqs.~(\ref{eq:ARP_lta_SG}) and (\ref{eq:ARP_lta_CG})
to the situation of a quantum walk over a ring with an odd number of nodes, for which $P_{\rm QW}^{\rm ring} = 1/N$. In the latter case 
there are no degenerate eigenvalues, so that $D(E_n)=1$ for all $n$. Note that from the value $P_{\rm QW}^{\rm ring} = 1/N$ one should not infer that the CTQW is never
faster than the CTRW. The result only implies that in the long-time limit the
return probability is in average equally distributed among all nodes of the network.

We now turn to the question in how far the situation changes when we (randomly) add bonds to the SG. 
By this we will explore the transition from the SG to the CG. As it will turn out, the CTQW spreading 
may increase when the disorder gets larger.

\section{From the star graph to the complete graph}

\begin{figure}[h]
 \centering
 \subfigure{\includegraphics[width=0.33\columnwidth]{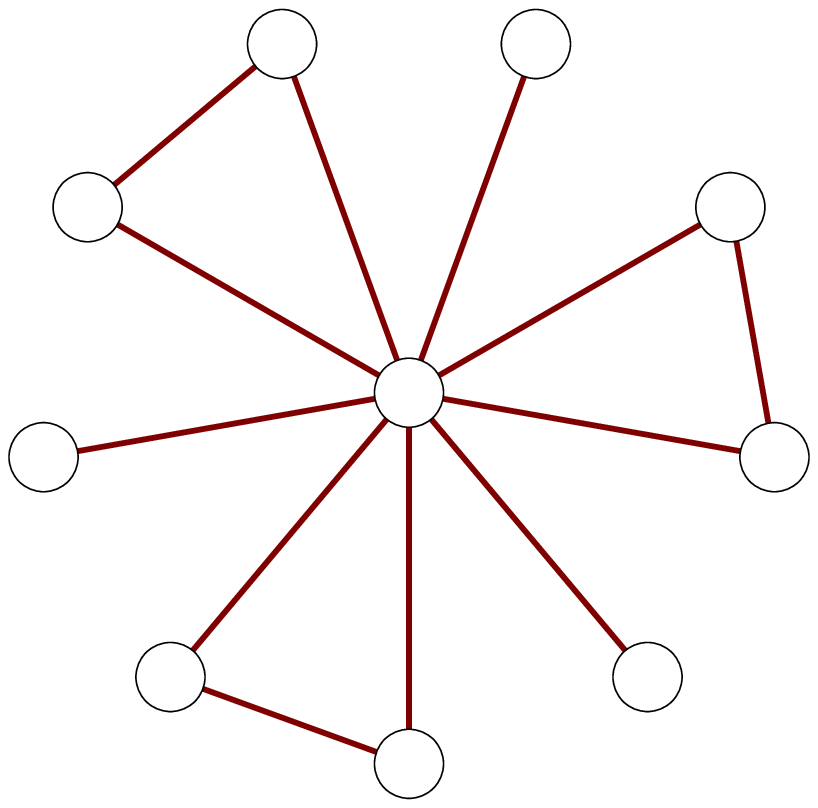}}
 \subfigure{\includegraphics[width=0.33\columnwidth]{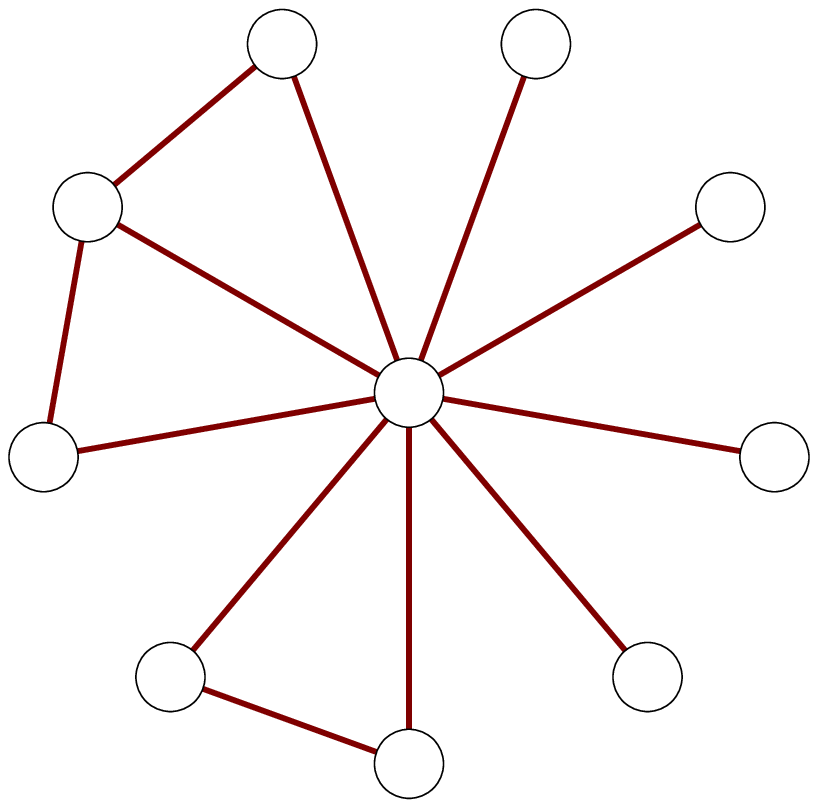}}
 \subfigure{\includegraphics[width=0.33\columnwidth]{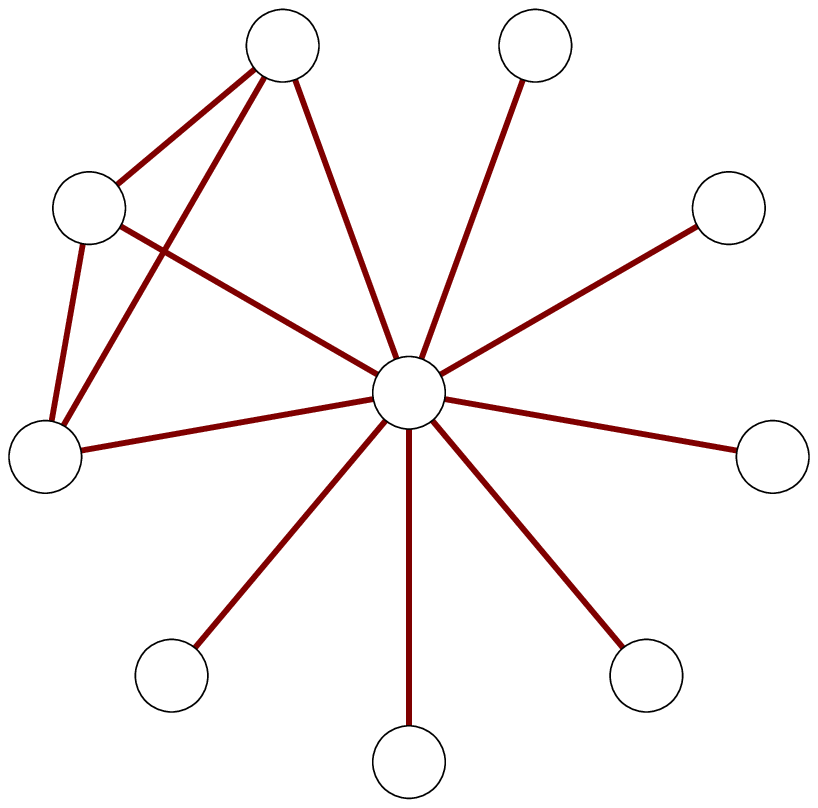}}
\caption[3 class]{Examples of SG with $B=3$ additional bonds}
\label{fig:1_2_classes}
\end{figure}

In this section we consider graphs generated from a SG of size $N$ to which $B$ additional bonds 
are added. The new bonds are not allowed to connect any node 
with itself; furthermore only a single bond between any pair of nodes is permitted. The total number of additional bonds, 
$B_{\rm max}(N)$, needed to convert the SG into the CG 
of the same size is:
\begin{equation}
  B_{\rm max}(N) =\frac{(N-1)(N-2)}{2}.
\end{equation}
Examples of SG with $B=3$ additional bonds are given in Figure \ref{fig:1_2_classes}. 

As seen from Fig.~\ref{fig:1_2_classes}, randomly adding bonds to a SG may lead to distinct 
topologies. While for $B$ very close to $1$ or to $N$ the number of topologically distinct graphs is small, this number 
increases rapidly when leaving these regions ($B/N\ll1$ and $(B-N)/N\ll1$). Now, each topological realization 
leads to a set of eigenvalues. Graphs with distinct topologies may have an identical set of eigenvalues 
(the graphs are then called co-spectral \cite{cvetkovic1980spectra}). For computational reasons we will 
not distinguish between such graphs and we will denote by the term "configuration" the set of graphs leading to the same set of 
eigenvalues.

We will call the set of all distinct configurations obtained from a SG
by adding $B$ additional bonds a ``clan'', and denote the number of elements in the clan by $\mathcal{N}_B$. 
Evidently, $\mathcal{N}_B$ depends on the size of the network. 
We determine for every fixed $B$ value the number of distinct eigenvalue sets inside the corresponding clan by 
considering all the corresponding Hamiltonians. This is done as follows: 
\begin{enumerate}
\item Generate the Hamiltonian $\bf {H}$ of the given SG (see Eqs.~(\ref{eqn:star_matrix})~and~(\ref{eq:star_hamiltonian})). There 
are $(N-1)(N-2)/2$ zero entries above (below) the diagonal (first row and column have only non-zero entries).
\item Consider all the possibilities of changing pairwise such entries from $\mathbf{H}_{ij}=\mathbf{H}_{ji}=0$ to 
$\mathbf{H}_{ij}=\mathbf{H}_{ji}=-1$, corresponding to the insertion of an additional bond between the nodes $i$ and $j$. 
There are $2^{\frac{(N-1)(N-2)}{2}}$ ways of doing this, leading to the same number of distinct $\bf H$.
\item For each $\bf {H}$ obtained in this way determine the eigenvalue set and sort the eigenvalues in ascending order. 
\item Group together all the $\bf H$ corresponding to a fixed $B$ value and determine by comparison of the ascending 
eigenvalues the number of distinct eigenvalue sets $\mathcal {N}_B$. 
\end{enumerate}

For the numerical procedures we use the standard package of Mathematica 8.0. To fix the ideas we start exemplarily 
with graphs of $N=10$ and with the corresponding clans. We are interested in how many configurations $\mathcal{N}_B$ are 
included in the clan $B$ and how many graphs are represented by a given configuration. For this we show in Table 
\ref{tab:LeastMostProbable} for several $B$ values the corresponding $\mathcal{N}_B$.

Moreover, it turns out that a configuration containing graphs which are less symmetric with respect to the exchange symmetry 
is more probable to show up when distributing the additional $B$ bonds randomly. To exemplify this we also show in 
Table \ref{tab:LeastMostProbable} the most probable and the least probable graph topology together 
with their eigenvalue sets. One remarkable feature of Table \ref{tab:LeastMostProbable} is that the least probable configurations 
have eigenvalue sets consisting of integer values only, while the 
eigenvalue sets corresponding to the most probable configurations 
also include non-integer polynomial roots.

\begin{table*}
\begin{tabularx}{\textwidth}{|l|l|c|c|X|X|}
\hline\hline
 &   & The least & The most & The least & The most \\
 $B$ & $\mathcal{N}_B$  &  probable  & probable & probable & probable \\
  &    & configuration & configuration & eigenvalue & eigenvalue \\
 &  & & & set & set \\
\hline\hline
\small 4 & \small 11 & 
\includegraphics[width=50pt]{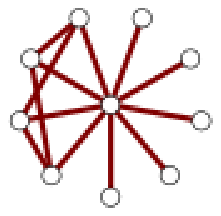} &
\includegraphics[width=50pt]{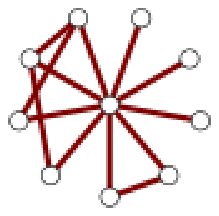}  &
\small \{10, 5, 3, 3, 1, 1, 1, 1, 1, 0\} & \small \{10, 4.414, 3, 3, 1.586,1, 1, 1, 1, 0\} \\
\hline
\small 10 & \small 1470 & 
\includegraphics[width=50pt]{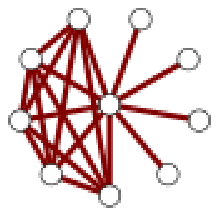} &
\includegraphics[width=50pt]{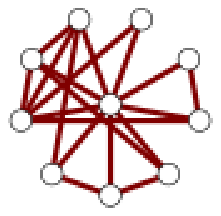} &
\small \{10, 6, 6, 6, 6, 1, 1, 1, 1, 0\} & \small \{10, 6.26, 5.4, 4.601, 3.4743, 2.625, 2.5858, 1.705, 1.34, 0\} \\
\hline
\small 18 & \small 31566 & 
\includegraphics[width=50pt]{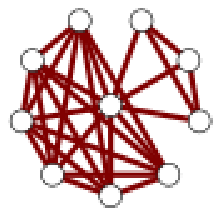} &
\includegraphics[width=50pt]{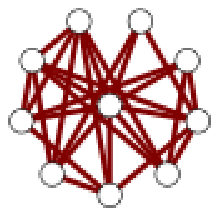} &
\small \{10, 7, 7, 7, 7, 7, 4, 4, 1, 0\} & \small \{10, 8.33, 7.5, 6.395, 5.6087, 5.3913, 4.605, 3.4562, 2.66, 0\} \\
\hline
\small 26 & \small 1470 & 
\includegraphics[width=50pt]{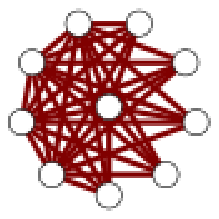} &
\includegraphics[width=50pt]{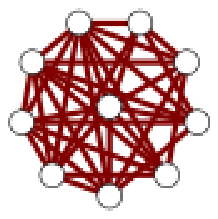} &
\small \{10, 10, 10, 10, 10, 5, 5, 5, 5, 0\} & \small \{10, 10, 9.408, 8.8192, 8.353, 7, 6.24, 5.396, 4.7835, 0\} \\
\hline
\small 32 & \small 11 & 
\includegraphics[width=50pt]{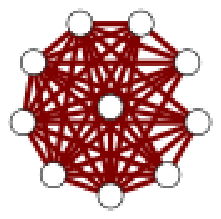} &
\includegraphics[width=50pt]{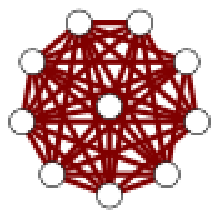} &
\small \{10, 10, 10, 10, 10, 10, 8, 8, 6, 0\} & \small \{10, 10, 10, 10, 10, 9.4142, 8, 8, 6.5858, 0\} \\
\hline

\end{tabularx}
\caption{For networks of size $N=10$ and different $B$ we display the least and the most probable eigenvalue sets.}
\label{tab:LeastMostProbable}
\end{table*}

\begin{figure}[ht!]
\centering
\includegraphics[width=\columnwidth]{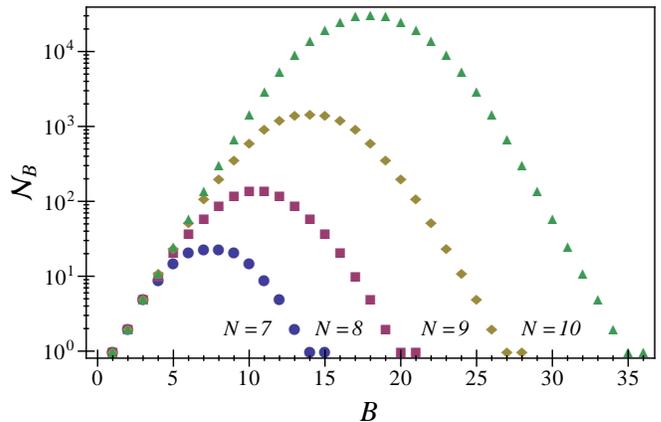}
\caption[Number of distinct eigenvalue sets]{Number of distinct eigenvalue
sets as a function of $B$ for the networks of $N=7$, $N=8$, $N=9$, and $N=10$ nodes. Note the logarithmic scale.}
\label{fig:Sets_number}
\end{figure}

Figure~\ref{fig:Sets_number} shows $\mathcal{N}_B$ as a function
of $B$ for different $N$. 
As can be seen from the figure, $\mathcal{N}_B$ reaches a maximum around $B(N)=B_{\rm max}(N)/2$. One may note the very 
large increase of $\mathcal{N}_B$ with growing $N$ (note the logarithmic scale on the $y$-axis). 
For $N=10$ (where $B_{max}=36$) and $B$ taken to be $B_{\rm max}/2=18$ 
we find more than $3\cdotp10^4$ distinct eigenvalue sets, while for $N=7$ (where $B_{max}=15$) and $B=8$ there are only $215$. For $N$ 
exceeding $10$ it gets quickly impossible to determine through the above algorithm all the configurations. Thus, for $N\geq10$ we will rely on 
randomly creating networks and obtain the ensemble average using a relatively large number $R=10000$ of realizations. Formally we set:
\begin{equation}
\label{eq:ensemble}
 \langle...\rangle_R\equiv\frac{1}{R}\sum_{r=1}^{R}[...]_r,
\end{equation}
where $r$ runs over the particular realizations. In this way we determine the ensemble-averaged 
probabilities $\langle\bar p(t)\rangle_R$ and
$\langle|\bar {\alpha}(t)|^2\rangle_R$, along with the long-time average
$\langle P_{\rm QW}\rangle_R$.

\subsection{Average probability of being at the origin}

The influence of additional bonds on the spreading is captured in the behavior of the transition probabilities
between certain pairs of nodes. However, as can be anticipated, the
functional form of these transition probabilities strongly depends on the
particular nodes chosen. Therefore, we start by considering the ensemble
averages of the average
probability to be (return or stay) at the initial node. In particular, we
focus on the ensemble average of $\bar p(t)$ for CTRW and of the lower bound $|\bar\alpha(t)|^2$ for
CTQW, see Eqs.~(\ref{eq:ARP}) and
(\ref{eq:ARP_LowerBound}).
Taking the time-average of $\langle|\bar\alpha(t)|^2\rangle_R$ yields $\langle P_{\rm
QW}\rangle_R$, see Eq.~(\ref{eq:ARP-QW-lta}).
Here and in the following we focus on graphs with $N=10$.

\begin{figure}[h]
 \centering
\includegraphics[width=\columnwidth]{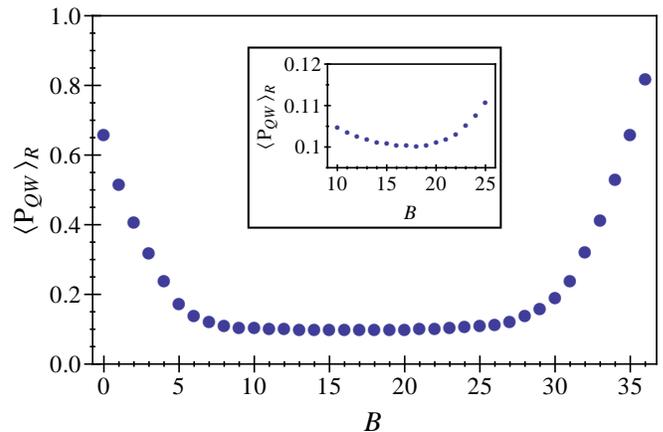}
\caption[lta]{$\langle P_{\rm QW}\rangle_R$ for networks of
$N=10$ nodes with $R=10000$.}
\label{fig:1_2_lta}
\end{figure}

Figure~\ref{fig:1_2_lta} shows $\langle P_{\rm
QW}\rangle_R$ as a function of the number of additional bonds. Clearly,
one observes a decrease from the SG-value to a broad minimal plateau
centered
around $B=B_{\rm max}/2$ (see the inset), after which the values increase again until the
CG-value is reached. 
The minimal value of $\langle P_{\rm QW}\rangle_R$ at $B=B_{\rm max}/2$
approaches $1/N$ from above. Therefore, when the number of additional
bonds is so large as to yield the maximal number of possible
configurations, the time-averaged spreading becomes comparable to that for
the corresponding CTRW. From this we can already infer that adding
(removing) bonds to (from) the SG (CG) enhances the spreading of CTQW. This behavior differs from the one found for
CTQW on SWN which are build up from a ring-like configuration; for SWN the spreading becomes 
slow with increasing $B$ \cite{mülken2007quantum}.

In order to highlight better the behavior of the
corresponding CTRW, we have to take a closer look at the temporal development
of the average probability to be at the initial node. For the SG and the
CG there exist exact analytic expressions for $|\bar {\alpha}(t)|^2$ and for $\bar p(t)$, 
see Eqs.~(\ref{eq:star_eff_clas}) and
(\ref{eq:compl_eff_quantum}). In both cases the lower bound
$|\bar {\alpha}(t)|^2$ shows oscillations but no decay to a value
comparable to $1/N$ (the long-time value for $\bar p(t)$). Specifically,
$|\bar {\alpha}(t)|^2$ oscillates around the values given by
Eqs.~(\ref{eq:ARP_lta_SG}) and (\ref{eq:ARP_lta_CG}) for the SG and for the CG, respectively.

\begin{figure}[ht!]
 \centering
 \subfigure{\includegraphics[scale=0.25]{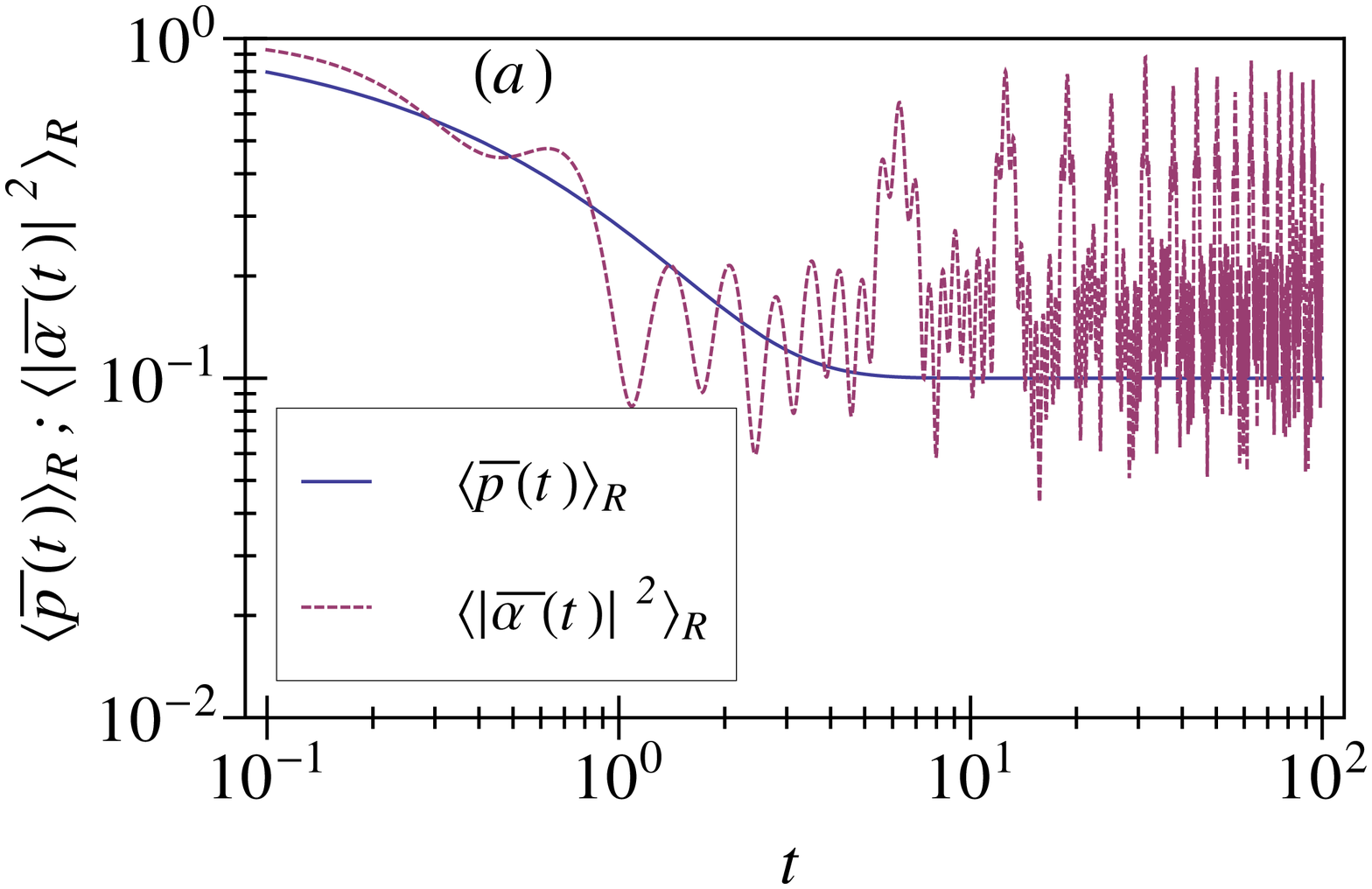}}
 \subfigure{\includegraphics[scale=0.25]{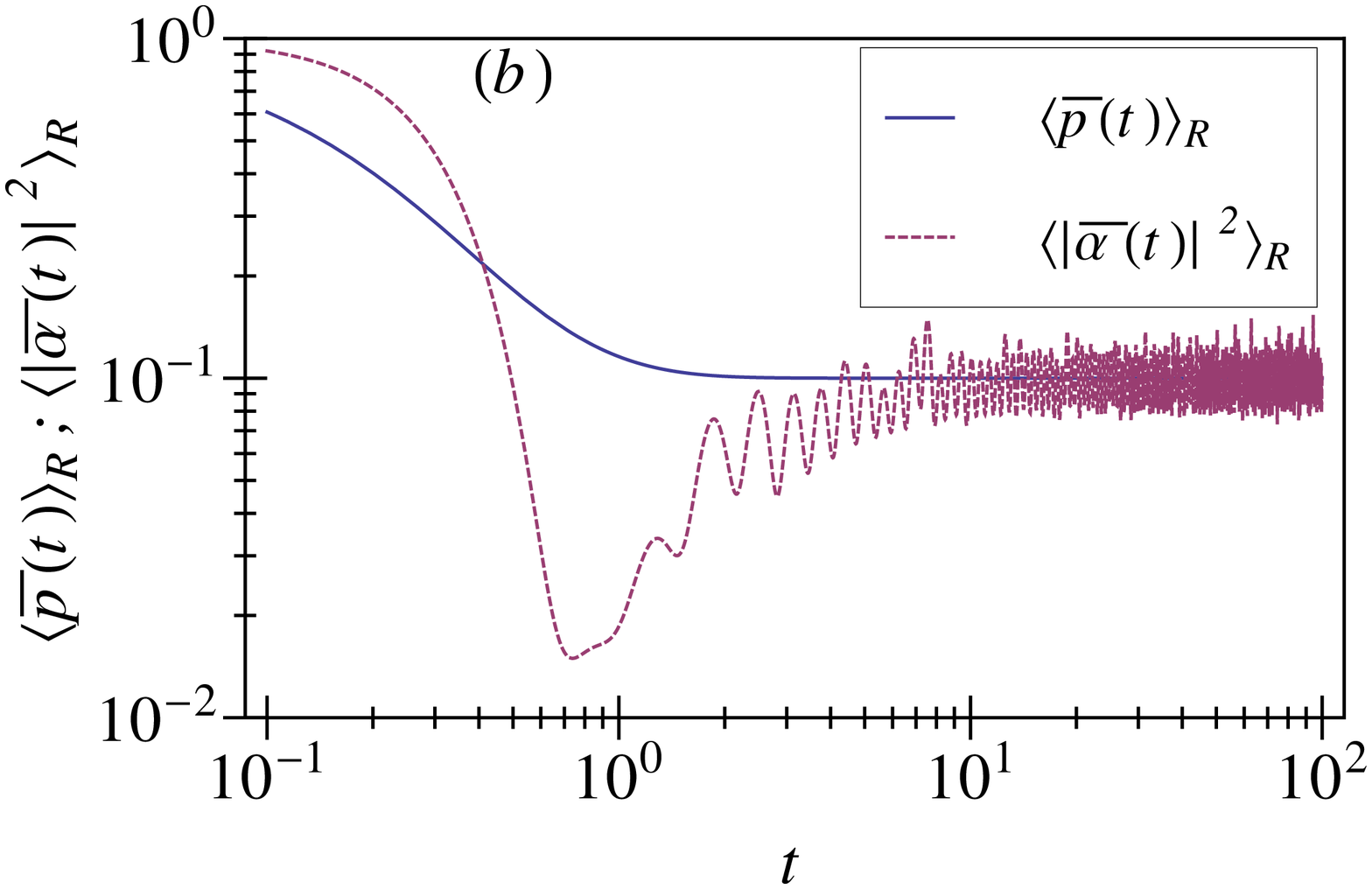}}
 \subfigure{\includegraphics[scale=0.25]{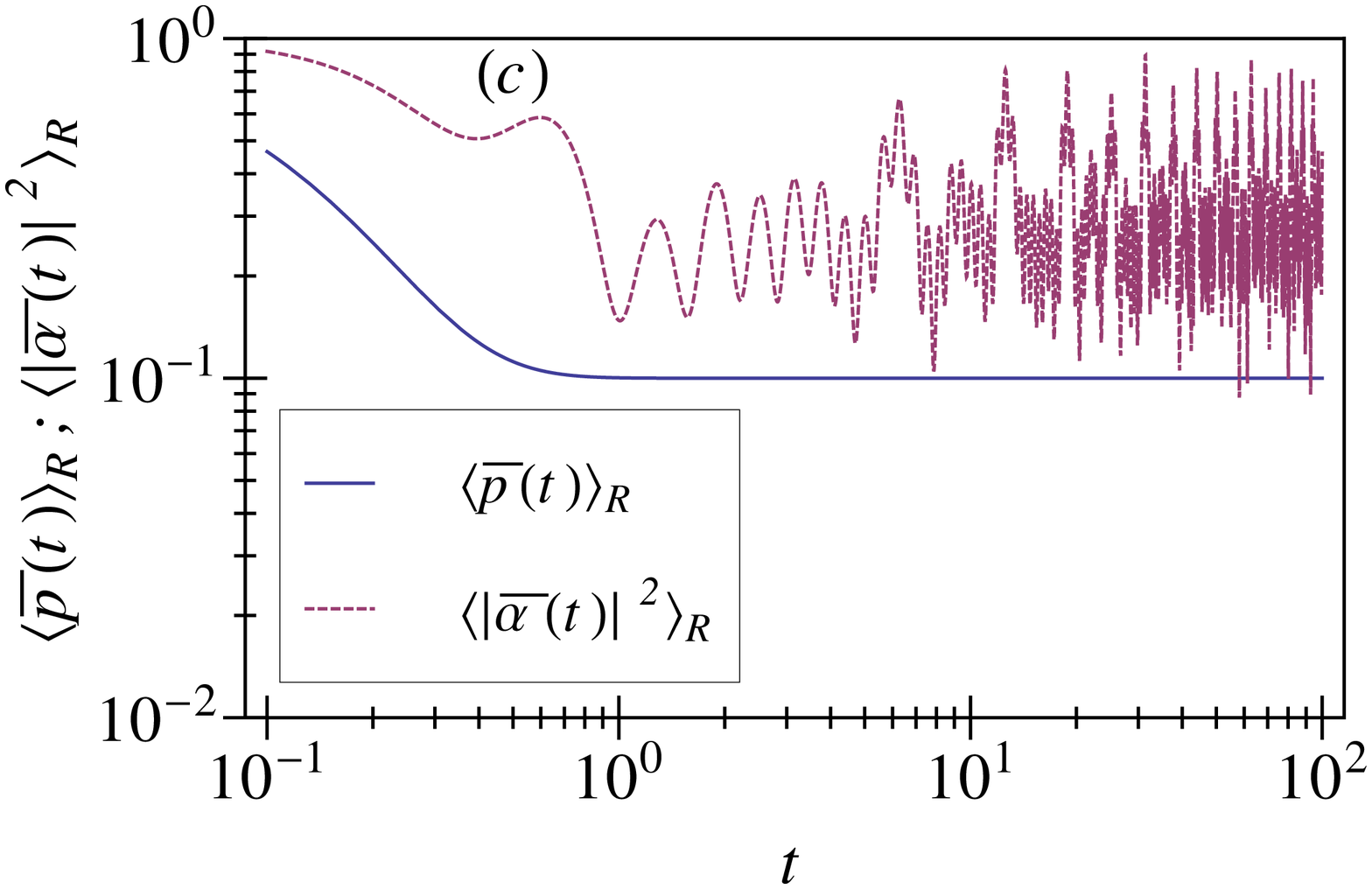}}
\caption[Ensemble]{Averaged probabilities
$\langle\bar p(t)\rangle_R$ and $\langle\bar \pi(t)\rangle_R$ on double logarithmic scales 
for the SG of size $N=10$ with $B=4$ (a), $18$ (b) and $32$ (c), where $R=10000$.}
\label{fig:Eff}
 \end{figure}

Now, on adding bonds to the SG $\langle|\bar{\alpha}(t)|^2\rangle_R$ 
follows a complex course: First it decreases to lower values until reaching $B=B_{\rm {max}}/2=18$; then it 
increases again towards the behavior found for the CG. This is exemplified in Fig.~\ref{fig:Eff} which shows $\langle|\bar
{\alpha}(t)|^2\rangle_R$ for networks of size $N=10$ with $B=4$, $18$, and
$32$ additional bonds. One notes that $\langle|\bar
{\alpha}(t)|^2\rangle_R$ for $B=4$ explores roughly the same interval as does $\langle \bar p(t)\rangle_R$,
see Fig.~\ref{fig:Eff}(a). Taking $B=B_{\rm max}/2=18$ yields a $\langle|\bar
{\alpha}(t)|^2\rangle_R$ whose time-average is close to $P_{RW}=1/N$ but which also - for short times - 
reaches values which are about one order of magnitude
below those of the CTRW, see Fig.~\ref{fig:Eff}(b). Adding more bonds
reverses the behavior, i.e., $\langle P_{\rm QW}\rangle_R$ becomes larger
and the amplitude of the oscillations around this value becomes smaller,
see Fig.~\ref{fig:Eff}(c).

Since $\langle|\bar{\alpha}(t)|^2\rangle_R$ is a lower bound, care has to
be taken when interpreting the above results. It still might be possible
that the exact value, namely $\langle\bar{\pi}(t)\rangle_R$, lies
above the CTRW curve. However, as has been shown earlier
\cite{mülken2006efficiency},
\cite{mulken2011continuous}, the maxima of
$\langle|\bar{\alpha}(t)|^2\rangle_R$ tend to be of similar value as the
maxima of $\langle\bar{\pi}(t)\rangle_R$. In particular, for
$B=B_{\rm max}/2$ there is a considerable short-time part (up to $t\approx 4$) where 
$\langle|\bar{\alpha}(t)|^2\rangle_R$ lies below
$\langle\bar{p}(t)\rangle_R$ and which includes several local maxima. Note that at $t\approx 4$
$\langle\bar{p}(t)\rangle_R$ has already reached the equipartition value.
We thus infer that in this case during the (short) time interval needed by
the CTRW to reach equipartition, the spreading of CTQW is faster than that
of the corresponding CTRW. 
\subsubsection*{Eigenvalue sets}

Both $\langle|\bar{\alpha}(t)|^2\rangle_R$ for CTQW as well as
$\langle\bar{p}(t)\rangle_R$ for CTRW depend only on the corresponding eigenvalues (of
the Hamiltonian and of the transfer matrix, respectively). As mentioned above, 
the SG and the CG have eigenvalue sets with one highly
degenerate eigenvalue. This situation will change as a function of $B$.

\begin{figure}[ht!]
\centering
\includegraphics[width=\columnwidth]{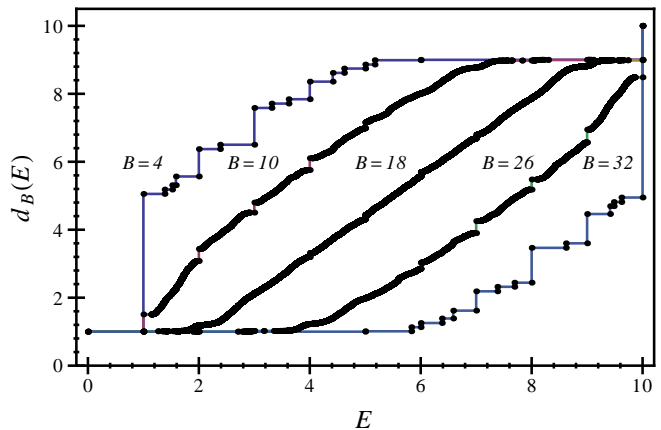}
\caption[Eigenvalue Average]{Aveage number of eigenvalues for $N=10$ below $E$; here the number of additional bonds is 
$B=4, 10, 18, 26$, and $32$, see text for details.}
\label{fig:EigenvalueAverage}
\end{figure}

In order to visualize the transition from the SG to the CG in the domain of the eigenvalues, 
we consider as a function of $B$ the quantity:
\begin{equation}
 d_B(E)=\langle \sum_{n}\theta(E-E_{n}^{(B)}) \rangle_R = \frac{1}{R}\sum_{n,r}\theta(E-[E_{n}^{(B)}]_r). 
\end{equation} 
Here $\theta(x)$ is the Heaviside function and $E_n^{(B)}$ are the eigenvalues of the $r$th realization of 
the distribution of the $B$ additional bonds. 
The quantity $d_B(E)$ gives the average number of eigenvalues below $E$. 
Figure~\ref{fig:EigenvalueAverage} shows $d_B(E)$ for $B=4,10,18,26$, and
$32$. 

With increasing $B$ the eigenvalues move to higher values, as is evident from the fact that the $d_B(E)$ curves decrease with 
increasing $B$. Sharp steps in $d_B(E)$ indicate eigenvalues with large degeneracy; such instances are particularly clear for $B=4$ 
and for $B=32$ whereas the curve for $B=18$ is much smoother. This is in line with the behavior of $|\bar{\alpha}(t)|^2$: a 
few highly degenerate eigenvalues lead to a plateau in $|\bar{\alpha}(t)|^2$ whereas non-degenerate 
eigenvalues let $|\bar{\alpha}(t)|^2$ drop to values close to the CTRW equipartition value $1/N$.

\section{Conclusion} \label{con}
We have studied the dynamics of continuous-time quantum walks on networks 
whose connectivity matrices have one highly degenerate eigenvalue. 
This fact leads to a slow spreading of CTQW in term of the average probability to return or remain at the initially excited node of the network. 
In particular, we focused on two types of networks, the star graph and the complete graph. We studied 
the crossover from the star graph to the complete graph by randomly adding bonds to the star graph until 
reaching the complete graph. In so doing we found that in the ensemble average the 
spreading gets faster when adding approximately $B=B_{\rm max}/2$ bonds to the star graph. 
The reason for this behavior is to be found in 
the ensemble averaged distribution of the eigenvalues of the connectivity matrices. Adding (extracting) 
more and more bonds to (from) the star (complete) graph results in broader distributions. Thus, we have
shown that under disorder, obtained by either adding bonds to the star graph or by deleting bonds from the complete 
graph, the quantum walks spread more.
 
\begin{acknowledgements}
Support from the Deutsche Forschungsgemeinschaft (DFG) and the Fonds der
Chemischen Industrie is gratefully acknowledged. We also thank Piet Schijven and Maxim Dolgushev for fruitful discussions.
\end{acknowledgements}


\bibliographystyle{spmpsci}

\begin{thebibliography}{10}
\providecommand{\url}[1]{{#1}}
\providecommand{\urlprefix}{URL }
\expandafter\ifx\csname urlstyle\endcsname\relax
  \providecommand{\doi}[1]{DOI~\discretionary{}{}{}#1}\else
  \providecommand{\doi}{DOI~\discretionary{}{}{}\begingroup
  \urlstyle{rm}\Url}\fi

\bibitem{alexander1982density}
Alexander, S., Orbach, R.: {Density of states on fractals: \textquotedblleft
  fractons\textquotedblright}.
\newblock J. Phys. (Paris) Lett. \textbf{43}, 625--631 (1982)

\bibitem{bray1988diffusion}
Bray, A., Rodgers, G.: {Diffusion in a sparsely connected space: A model for
  glassy relaxation}.
\newblock Phys. Rev. B \textbf{38}, 11461 (1988)

\bibitem{burgarth2008coupling}
Burgarth, D., Maruyama, K., Nori, F.: Coupling strength estimation for spin
  chains despite restricted access.
\newblock Arxiv preprint arXiv:0810.2866  (2008)

\bibitem{cvetkovic1980spectra}
Cvetkovi\'c, D., Doob, M., Sachs, H.: Spectra of Graphs: Theory and
  Applications.
\newblock Academic Press, New York (1997)

\bibitem{darázs2010pólya}
Dar{\'a}zs, Z., Kiss, T.: P{\'o}lya number of the continuous-time quantum
  walks.
\newblock Phys. Rev. A \textbf{81}, 062319 (2010)

\bibitem{doi1988theory}
Doi, M., Edwards, S.: {The theory of polymer dynamics}.
\newblock Oxford University Press, Oxford (1988)

\bibitem{farhi1998quantum}
Farhi, E., Gutmann, S.: {Quantum computation and decision trees}.
\newblock Phys. Rev. A \textbf{58}, 915 (1998)

\bibitem{van1992stochastic}
van Kampen, N.: {Stochastic processes in physics and chemistry}.
\newblock Amsterdam: North Holland (1992)

\bibitem{JKempe}
Kempe, J.: {Quantum random walks - an introductory overview}.
\newblock Contemporary Physics \textbf{44}, 307 (2003)

\bibitem{kenkre1982exciton}
Kenkre, V., Reineker, P.: {Exciton dynamics in molecular crystals and
  aggregates}.
\newblock Springer, Berlin (1982)

\bibitem{machida2011discrete}
Machida, T.: {The discrete-time Grover walk on star graphs with one loop}.
\newblock Arxiv preprint arXiv:1103.1280  (2011)

\bibitem{mülken2006coherent}
M{\"u}lken, O., Bierbaum, V., Blumen, A.: {Coherent exciton transport in
  dendrimers and continuous-time quantum walks}.
\newblock J. Chem. Phys. \textbf{124}, 124905 (2006)

\bibitem{mülken2005spacetime}
M{\"u}lken, O., Blumen, A.: {Spacetime structures of continuous-time quantum
  walks}.
\newblock Phys. Rev. E \textbf{71}, 036128 (2005)

\bibitem{mülken2006efficiency}
M{\"u}lken, O., Blumen, A.: {Efficiency of quantum and classical transport on
  graphs}.
\newblock Phys. Rev. E \textbf{73}, 066117 (2006)

\bibitem{mulken2011continuous}
M{\"u}lken, O., Blumen, A.: Continuous-time quantum walks: Models for coherent
  transport on complex networks.
\newblock Phys. Rep. \textbf{502}, 37 (2011)

\bibitem{mülken2007quantum}
M{\"u}lken, O., Pernice, V., Blumen, A.: {Quantum transport on small-world
  networks: A continuous-time quantum walk approach}.
\newblock Phys. Rev. E \textbf{76}, 051125 (2007)

\bibitem{mülken2005asymmetries}
M{\"u}lken, O., Volta, A., Blumen, A.: {Asymmetries in symmetric quantum walks
  on two-dimensional networks}.
\newblock Phys. Rev. A \textbf{72}, 042334 (2005)

\bibitem{PhysRevA.79.012323}
Reitzner, D., Hillery, M., Feldman, E., Bu\ifmmode~\check{z}\else \v{z}\fi{}ek,
  V.: Quantum searches on highly symmetric graphs.
\newblock Phys. Rev. A \textbf{79}, 012323 (2009)

\bibitem{salimi2009continuous}
Salimi, S.: Continuous-time quantum walks on star graphs.
\newblock Ann. Phys. \textbf{324}, 1185--1193 (2009)

\bibitem{xu2009exact}
Xu, X.: {Exact analytical results for quantum walks on star graphs}.
\newblock J. Phys. A \textbf{42}, 115205 (2009)

\bibitem{ziman1979principles}
Ziman, J.: Principles of the Theory of Solids.
\newblock Cambridge University Press, Cambridge, England (1979)

\end{thebibliography}

%
%

\end{document}